\documentclass[aps,prc,preprint,showpacs,floatfix,nofootinbib]{revtex4}

\usepackage{amsmath}
\usepackage{amssymb}
\newcommand{\be}{\begin{equation}}
\newcommand{\ee}{\end{equation}}
\newcommand{\bel}[1]{\be\label{#1}}
\newcommand{\re}[1]{Eq.~(\ref{#1})}

\newcommand{\hsp}{\hspace*{1pt}}
\newcommand{\hspm}{\hspace*{.5pt}}
\newcommand{\ov}[1]{\overline{#1}}
\usepackage{graphicx}
\usepackage{bm}
\usepackage{color}

\begin{document}
\title{Entropy production in chemically nonequilibrium quark-gluon \\plasma created in central Pb+Pb collisions
at energies available at the CERN Large Hadron Collider}

\author{V.~Vovchenko$^{1,2,3}$, M.I.~Gorenstein$^{1,4}$, L.M.~Satarov$^{1,5}$,\\
I.N.~Mishustin$^{1,5}$, L.P.~Csernai$^6$, I.~Kisel$^{1,2}$ and  H.~St\"ocker$^{1,2,7}$}
\affiliation{
$^1$\mbox{Frankfurt Institute for Advanced Studies, D-60438 Frankfurt am Main, Germany}\\
$^2$\mbox{Johann Wolfgang Goethe Universit\"at, D-60438 Frankfurt am Main, Germany}\\
$^3$\mbox{Taras Shevchenko National University of Kiev, 03022 Kiev, Ukraine}\\
$^4$\mbox{Bogolyubov Institute for Theoretical Physics, 03680 Kiev, Ukraine}\\
$^5$\mbox{National Research Center ''Kurchatov Institute'', 123182 Moscow, Russia}\\
$^6$\mbox{Institute for Physics and Technology, University of Bergen, 5007 Bergen, Norway}\\
$^7$\mbox{GSI Helmholtzzentrum f\"ur Schwerionenforschung GmbH, D-64291 Darmstadt, Germany}\\
}

\begin{abstract}
We study the possibility that partonic matter produced
at an early stage of ultrarelativistic
heavy-ion collisions is out of chemical equilibrium.
It is assumed that initially this matter is mostly composed of gluons,
but quarks and antiquarks are produced at later times.
The dynamical evolution of partonic system is described by the
Bjorken-like ideal hydrodynamics with a time dependent quark fugacity.
The results of this model are compared with those  obtained by assuming
the complete chemical equilibrium of partons  already at the initial stage.
It is shown that in a~chemically non-equilibrium scenario the entropy
gradually increases, and about 25\%
of the total final entropy is generated during the hydrodynamic evolution
 of deconfined matter. We argue that the (anti)quark suppression included
 in this approach may be responsible for reduced (anti)baryon to meson ratios
 observed in heavy-ion collisions at energies available at the CERN Large Hadron Collider.
\end{abstract}

\maketitle

\section{Introduction}

Relativistic heavy-ion collisions open the possibility to create in the laboratory strongly interacting matter under extreme conditions of high excitation energies and particle densities. One of the central questions is how the initial highly nonequilibrated
system evolves to a state of partial thermodynamic equilibrium. There exists several models which describe the initial state in terms of non-equilibrium parton cascades~\cite{Gyulassy,XuGreiner}, minijets~\cite{Eskola}, color glass condensate~\cite{McLerran}, coherent chromofields~\cite{magas-csernai,mishustin-kapusta} etc.

Relatively large gluon-gluon cross sections lead to the idea \cite{Pok-Hove}
that the gluonic components of colliding nucleons interact more strongly than the quark-antiquark ones.
As demonstrated in Ref.~\cite{Blaizot}, strong non-equilibrium effects in the gluonic sector persist only for a short
time~$\sim 1/Q_s$, where $Q_s\simeq 1-2~\textrm{GeV}$ is the so-called saturation scale~\cite{Gribov}, but at later times the system reaches a state of a partial thermodynamic equilibrium.
The two-step equilibration scenario of the quark-gluon plasma (QGP) was proposed~in \cite{raha,shuryak,sinha}.
It was assumed that the gluon thermalization takes place at the proper time $\tau_g<1~\textrm{fm}/c$ and the (anti)quarks equilibration occurs at $\tau_{\rm th}>\tau_g$. The estimates of Ref.~\cite{XuGreiner} show that $\tau_{\rm th}$ can be
of the order of~$5~\textrm{fm}/c$\hsp . Later,
such a scenario for heavy-ion collisions was considered by several authors, see
e.g.~\mbox{\cite{Bir93,Stri94,Tra96,Ell00,Dut02,Gel04,Scardina2013,Liu14,Mon14,Ruggieri2015}}. Recently the {\it pure glue} scenario for the initial state of Pb+Pb collisions at Relativistic Heavy Ion Collider (RHIC) and Large Hadron Collider~(LHC) energies was proposed in~\mbox{\cite{Sto16,Sto15b}}.

In this paper we describe the evolution of QGP produced in central heavy-ion collisions
by using one-dimensional scaling hydrodynamics.
In addition to the chemically equilibrated system we also consider
a pure glue initial scenario,  in which the QGP contains no quarks and antiquarks at the initial state
of its evolution.
 Below we introduce the  effective number of quark degrees of freedom and study the
sensitivity of system evolution to the chemical equilibration  time.
Special attention is paid to the entropy production in  this chemically
non-equilibrium scenario.  It is commonly accepted that an additional entropy can be created due to
dissipative processes  which are usually described in the framework of viscous hydrodynamics.
In the present work, a different mechanism of entropy production is investigated:
 we show that it may increase  during chemically non-equilibrium expansion
 of matter even in the ideal hydrodynamics.
 Earlier the role of chemically nonequilibrium effects in entropy evolution of purely hadronic systems
 was considered in Refs.~\cite{Beb92,Pra14}, but without a quantitative analysis of the total entropy
change.

The paper is organized as follows: In Sec.~II we study thermodynamic functions
of a~chemically undersaturated QGP (uQGP) at different temperatures and quark fugacities.
We obtain explicit relations for relative entropy growth in different scenarios of
the system evolution. In Sec.~III we formulate a~simplified model for describing the
hydrodynamic evolution of uQGP in heavy-ion collisions.
In Sec. IV we present our numerical results and analyze their
sensitivity to chemical equilibration time. The summary and outlook
are given in Sec.~V. Some preliminary results of this paper were presented in Ref.~\cite{Sto15b}.

\section{Thermodynamics of chemically under\-satu\-rated QGP}

Below we describe the QGP matter produced in heavy-ion collisions by the equation of state (EoS)
of an ideal gluon-quark-antiquark gas.
It is assumed that gluons are in full thermodynamic equilibrium while quarks and anti-quarks are in thermal equilibrium, but not necessarily in chemical equilibrium.
In this section the thermodynamic functions of chemically nonequilibrium QGP are obtained, and the chemical equilibration process in a static box is investigated.
\subsection{Thermodynamic functions of uQGP}
In the following we consider systems with equal numbers of quarks and antiquarks.
In the pure glue initial scenario there is undersaturation of (anti)quarks, hence, their chemical potentials are negative:
\bel{cemp}
\mu_q=\mu_{\ov{q}}\equiv\mu<0\,.
\ee
We define the (anti)quark fugacity as
\bel{cemp1}
\lambda=e^{\hsp\mu/T}<1\,.
\ee
The phase-space distribution functions of the ideal gas of massless quarks and antiquarks can be written as\hsp\footnote
{
Units $\hbar=c=k=1$ are used throughout the paper.
}
\bel{qdf}
f_q(\bm{p})=f_{\ov{q}}(\bm{p})=\frac{g_q}{(2\pi)^3}
\left[\exp{\left(\frac{p-\mu}{T}\right)}+1\right]^{-1}=
\frac{g_q\lambda}{(2\pi)^3}\left[\exp{\left(\frac{p}{T}\right)}+\lambda\right]^{-1}\,,
\ee
where $\bm p$ is the (anti)quark three--momentum in the fluid's rest frame and the degeneracy factor
$g_q=2\hspm N_c\hspm N_f$, where $N_c=3$ is the number of colors, and $N_f$ is the number of quark flavours.
Unless stated otherwise, we assume that $N_f=3$.

\begin{figure}[ht]
\centering
\includegraphics[width=0.75\textwidth]{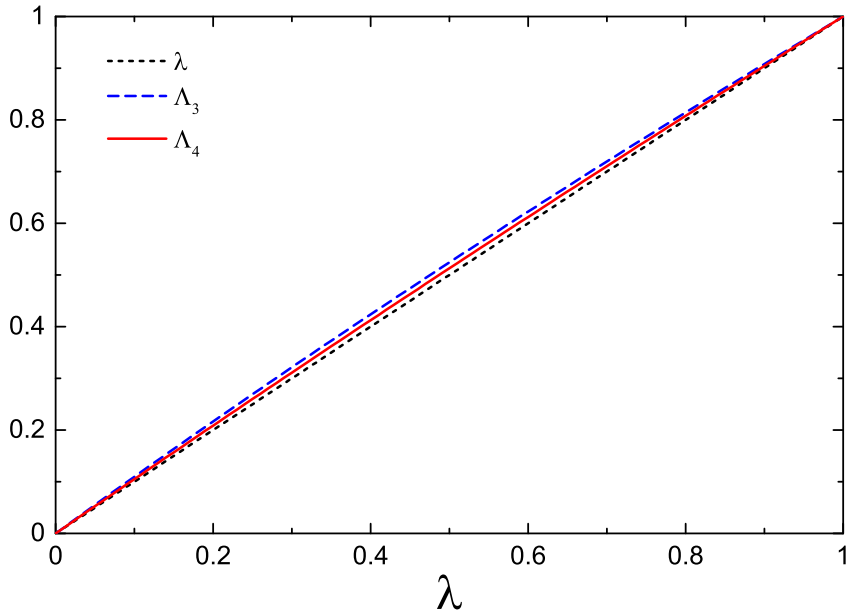}
\caption[]{
The functions $\Lambda_3$ and $\Lambda_4$ versus the quark fugacity $\lambda$.
}\label{fig:La}
\end{figure}

The Fermi-Dirac integral of the $n$-th order is defined as
\bel{fdi}
\varphi_{\hsp n} (\lambda)=\frac{\lambda}{\Gamma (n)}\int\limits_0^\infty \frac{dx x^{n-1}}
{e^{x}+\lambda}=\sum\limits_{k=1}^{\infty}(-1)^{k+1}\lambda^k\hsp k^{-n}\,.
\ee
It is easy to show that \mbox{$\varphi^{\,\prime}_n=\varphi_{\hsp n-1}/\lambda$}, and therefore,
$\varphi_{n}(\lambda)$ monotonically increases in the inter\-val~$\lambda\in[0;1]$.
Instead of $\varphi_{\hsp n}$ it is useful to introduce the function
\bel{fdi4}
\Lambda_{\hspm n}(\lambda)\equiv\frac{\varphi_{\hsp n} (\lambda)}{\varphi_{\hsp n} (1)}=
\frac{\lambda-\lambda^2\hsp 2^{-n}+\lambda^3\hsp 3^{-n}-\ldots}{1-2^{-n}+3^{-n}-\ldots}~,
\ee
which is normalized to unity at $\lambda=1$. The functions $\lambda$, $\Lambda_3$, and $\Lambda_4$ are
compared in Fig.~\ref{fig:La}. It is seen that they are very close to each other. Thus, one can safely
use the approximate relations $\Lambda_4 \simeq \lambda$ and $\Lambda_3 \simeq \lambda$\hspm .

Using~\re{qdf} one can represent the partial energy density $\varepsilon_q$ and pressure $P_q$
of quarks and antiquarks as functions of $T$ and $\mu$\hsp :
\bel{enpq}
\varepsilon_q=3\hspm P_q=\int d^{\hsp 3}p\left(f_q+f_{\ov{q}}\right)p=
\frac{\lambda\hsp g_q\hsp}{\pi^2}\int\limits_{0}^{\infty}
dp\hsp p^{\hsp 3}\left(e^{p/T}+\lambda\right)^{-1}=\frac{6\hspm g_q}{\pi^2}T^4\varphi_4(\lambda)=
3P_q^{\rm\, eq}(T)\hspm \Lambda_4(\lambda)\,,
\ee
where
\bel{enpq1}
P_q^{\rm\, eq}(T)=\frac{2\hspm g_q}{\pi^2}\,T^4\varphi_4(1)= {\frac{7\pi^2}{6\hspm 0}\,N_f}T^4
\ee
is the chemically equilibrated value of the (anti)quark pressure at~\mbox{$\lambda=1$}\,. Here and below we use the superscript
'eq'~to mark characteristics of chemically equilibrated matter.

By using Eqs.~(\ref{enpq})
one can calculate the total density of quarks and antiquarks
\bel{dnq1}
n_q=\left(\frac{\partial P_q}{\partial\mu}\right)_T=\frac{2\hspm g_q}{\pi^2}\,T^{\hsp 3}\varphi_{\hsp 3}(\lambda)=
n_q^{\rm eq}(T)\hsp\Lambda_3(\lambda)\,,
\ee
where
\bel{dneq}
n_q^{\rm eq}(T)=\frac{2\hspm g_qT^{\hsp 3}\hspace*{-1mm}}{\pi^2}\,\varphi_{\hsp 3}(1)
\ee
is the chemically equilibrated value of the (anti)quark density. The above equations show
that~$\Lambda_3$ and $\Lambda_4$ are, respectively, the suppression factors of density and  energy density
of (anti)quarks in the hot glue initial scenario as compared to the equilibrium case\hsp\footnote
{
 As follows from the relations $\Lambda_3\simeq\Lambda_4\simeq\lambda$,
both these suppression factors are approximately equal to the~{quark} fugacity $\lambda$.
}.
Using~\re{cemp1} one can evaluate the contribution of (anti)quarks to the entropy density,\hsp\footnote
{
The same relation follows from the thermodynamic identity $Ts_q=\varepsilon_q+P_q-\mu\hspm n_q$.
}
\bel{entrd}
s_q=\left(\frac{\partial P_q}{\partial T}\right)_\mu=\frac{4P_q}{T}+n_q\ln{\left(\lambda^{-1}\right)}.
\ee

Neglecting deviations from chemical equilibrium for gluons  we get the following relations for
gluonic parts of energy density $\varepsilon_g$, pressure $P_g$, and entropy density $s_g$ :
\bel{gltf}
\varepsilon_g=3P_g=\frac{3}{4}\,s_gT=\frac{8\hsp\pi^2}{15}\,T^4\,.
\ee
Adding the contributions of gluons, quarks and antiquarks gives the expressions
for the total energy density $\varepsilon$, pressure $P$, and entropy density $s$ density of the uQGP:
\begin{eqnarray}
&&\varepsilon=3P=\frac{8\hspm\pi^2}{15}\hsp T^4\Big[1+\alpha\Lambda_4(\lambda)\hsp\Big],\label{toten}\\
&&s=\frac{32\hsp\pi^2}{45}\hsp T^3\Big[1+\alpha\Lambda_4(\lambda)-
\beta\Lambda_3(\lambda)\ln{\lambda}\hsp\Big],\label{tots}
\end{eqnarray}
where
\bel{con1}
\alpha=\frac{7 g_q}{64\,}\simeq  {0.656\hsp N_f}\hsp ,~~~\beta=\frac{45\hsp g_q}{16\hsp\pi^4}\,\varphi_3(1)\simeq {0.156\hsp N_f}\hsp .
\ee
In the last equality of \re{con1} we use the relation $\varphi_3(1)=3\hsp\hspm\xi(3)/4$, where $\xi(3)\simeq 1.20\hspm 2$
is the Riemann zeta function $\xi(x)=\sum\limits_{k=1}^\infty k^{-x}$ at $x=3$\hsp .
\subsection{Equilibration in a box}
 Let us consider first the evolution of a homogeneous, chemically nonequilibrium QGP in a~static box of the volume $V$.
We assume that initially this plasma contains only gluons (i.e.~$\lambda=0$) at the temperature $T=T_0$\,. In absence of
partons' exchange with the box exterior, the system should approach the equilibrium state with $\lambda=1$ at large times.
In general case, the energy- and entropy densities of the system in intermediate states are functions of both $T$ and $\lambda$\,.
The time evolution of temperature depends on boundary conditions which in turn determine the type of a thermodynamic process.
We consider two limiting cases: the isothermal process ($T=T_0$), which requires some heat transfer from outside, and the process
with fixed energy ($\varepsilon=\varepsilon_0$). The second case corresponds to a thermally isolated system without any heat
exchange\hsp\footnote
{
 It will be shown that the relative change of the total entropy as a function of $\lambda$ coincides
in this case with the corresponding quantity for a Bjorken-like expanding QGP.
}.

\begin{figure}[htb]
\centering
\includegraphics[trim=0 7.5cm 0 8.5cm, clip, width=0.8\textwidth]{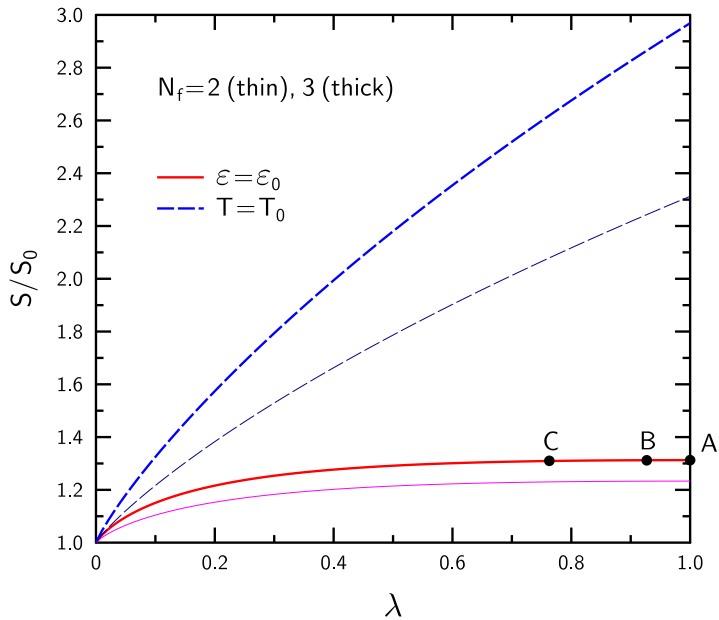}
\caption[]{
Relative increase of entropy of uQGP with respect to the pure glue initial state as a function of quark fugacity $\lambda$. Solid (dashed) lines correspond to box calculations with constant energy density (temperature). Thin and thick lines are calculated for the quark flavor numbers 2 and 3, respectively. Dots correspond to freeze-out states, estimated from Bjorken hydrodynamic analysis of central Pb+Pb collisions at the LHC energy (see below).
}\label{fig:entref1}
\end{figure}

 From~\re{tots} one gets the expression for the ratio of the total entropy $S=sV$ with respect to its initial value:
\bel{rtent}
\frac{S}{S_0}=\left(\frac{T}{T_0}\right)^3\big(1+\alpha\Lambda_4-\beta\Lambda_3\ln{\lambda}\big)\hsp.
\ee
Here and below we omit arguments $\lambda$ in the functions $\Lambda_3$ and $\Lambda_4$\hsp . In the isothermal case
we obtain
\bel{rtent1}
\frac{S}{S_0}=1+\alpha\Lambda_4-\beta\Lambda_3\ln{\lambda}~
\underset{\small\lambda\hsp\to 1}\longrightarrow ~1+\alpha~~~~~~(T=T_0)\,.
\ee
Note that the gluon fraction of entropy (the first term in the right hand side) does not change with time
in the isothermal process. One can see that the relative increase of entropy is proportional to the number
of quark flavors $N_f$\hsp.

\begin{figure}[htb]
\centering
\includegraphics[trim=0 7.5cm 0 8.5cm, clip, width=0.8\textwidth]{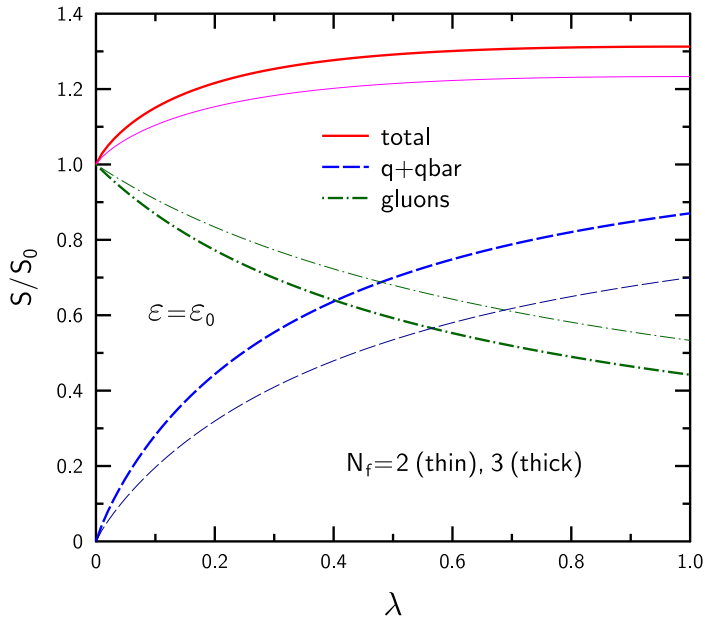}
\caption[]{
Entropy ratios in uQGP with respect to the pure glue initial state as functions of quark fugacity $\lambda$. Thin and thick lines correspond to box calculations with constant energy density for the quark flavor numbers 2 and 3, respectively. The dashed (dash-dotted) lines give contributions of $q\ov{q}$ pairs (gluons) to the total entropy ratios (the solid curves).
}\label{fig:entref2}
\end{figure}

The fixed energy case is more complicated. As one can see from~\re{toten}, the systems cools down during the process
of chemical equilibration:
\bel{temfe}
T=T_0\,\big(1+\alpha\Lambda_4)^{-1/4}~~~~~~(\varepsilon=\varepsilon_0)\,.
\ee
Substituting this temperature into~\re{rtent} gives the following result
\bel{ss0ef}
\frac{S}{S_0}=\frac{1+\alpha\Lambda_4-\beta\Lambda_3\ln{\lambda}}{(1+\alpha\Lambda_4)^{\hsp 3/4}}
\underset{\small\lambda\hsp\to 1}\longrightarrow ~(1+\alpha)^{1/4}~~~~~~(\varepsilon=\varepsilon_0)\,.
\ee
From the comparison of Eqs.~(\ref{rtent1} and (\ref{ss0ef}) one can see that the relative increase of entropy
is smaller in the second case. The asymptotic values of $S/S_0$ are equal in this case to 23\% and~31\%  (approximately)
for $N_f=2$ and 3, respectively.

 Figure~\ref{fig:entref1} shows the results of numerical calculations of $S/S_0$ for both considered cases.
In Fig.~\ref{fig:entref2} we consider in more details the system equilibration at $\varepsilon=\varepsilon_0$\hsp .
One can see that in this case the fraction of entropy contained in gluons decreases with time, however this decrease is more than
compensated by a rising contribution of $q\ov{q}$ pairs.
\section{Evolution of undersaturated QGP within the Bjorken \mbox{hydrodynamics}}
We consider central Pb+Pb collisions at ultrarelativistic energies
of the LHC. Our calculations below are performed under the following assumptions:

\noindent\textit{Equation of state}\\
The matter produced in the central rapidity region at LHC energies
has a nearly vanishing net baryon density.
To describe this matter we apply the EoS of an ideal gas
of massless gluons, quarks and antiquarks
obtained in Sec.~IIA.
According to~\re{toten}, this EoS can be written in the Stefan--Boltzmann
form, $\varepsilon=3P=\sigma\hsp T^4$, where the
coefficient $\sigma\propto 1+\alpha\Lambda_4(\lambda)$.
The first and second terms in this expression describe, respectively,
the contributions of gluons\hsp\footnote
{
As already mentioned, we neglect deviations from chemical equilibrium
for gluons during the whole process of the uQGP evolution.
}
and $q\ov{q}$ pairs.
The quantity~$\Lambda_4$ is the quark suppression factor, approximately
equal to the (anti)quark fugacity~$\lambda$.

\noindent\textit{Bjorken hydrodynamics}\\
Space-time evolution of uQGP is described by
the ideal relativistic hydrodynamics. Corres\-ponding equations of motion can be written as
\bel{hydro}
\frac{\partial T^{\mu\nu}}{\partial x^{\nu}}~=~0\,,
\ee
where
\bel{tmunu}
T^{\mu\nu}~=~(\varepsilon + P)\hsp u^{\mu}u^{\nu}~-~Pg^{\mu\nu}~
\ee
is the energy-momentum tensor, $u^\mu$ is the flow four-velocity, and $g^{\mu\nu}$ is the
diagonal metric tensor with $g^{00}=-g^{11}=-g^{22}=-g^{33}=1$.

Below we neglect the transverse motion of matter created in a nuclear collision.
The center of mass frame will be used with the longitudinal axis $z$ taken along the beam direction.
Following the Bjorken model~\cite{bjorken} we assume that a thermally  (but not necessary chemically) equilibrated QGP has been created at $\tau=\tau_0,\,r_\perp=\sqrt{x^2+y^2}<R_A$, where
$\tau=\sqrt{t^2-z^2}$ is the proper time of a fluid element and~$R_A$ is the geometrical radius of initial nuclei.
We consider only the (1+1) dimensional,
boost-invariant solution of hydrodynamic equations which satisfies the conditions \cite{bjorken,hwa,hwa1,chiu,chiu1,goren}:
\bel{boost}
u^{\mu}=\frac{1}{\tau}(t,{\bf 0},z)^\mu,~~~
\varepsilon=\varepsilon (\tau)\,,~~~~~~P=P(\tau)\,.
\ee
Using these relations one can show that~Eqs. (\ref{hydro}) and (\ref{tmunu}) are reduced to
the equation
\bel{eq-eps}
\frac{d\hsp\varepsilon}{d\hspm\tau}+\frac{\varepsilon +P}{\tau}=0\,.
\ee
Substituting the relation $P=\varepsilon/3$ in Eq.~(\ref{eq-eps}) one obtains:
\bel{eps-tau}
\varepsilon~=~\varepsilon(\tau_0)\,\left(\frac{\tau_0}{\tau}\right)^{4/3},
\ee
where the parameter $\tau_0$ corresponds to the initial proper time of the hydrodynamic expansion.

\noindent\textit{Entropy increase}\\
For the boost-invariant Bjorken expansion,
the total entropy per unit space-time rapidity can be expressed as~\cite{Sat07}
\bel{eq:dSdy}
\frac{dS(\tau)}{d \eta} =\pi \, R^{\hsp 2}_A \, s(\tau) \, \tau\,,
\ee
where the space-time rapidity $\eta$ is defined as $\eta=\tanh^{-1}(z/t)$\hsp .
Because of the boost invariance, $dS/d\eta$
does not depend on $\eta$ within the Bjorken model.
In the case of chemical equilibrium, i.e., when $\lambda=1$,
the entropy density  (in the net baryon-free matter)
$s=(\varepsilon+P)/T$ is inversely proportional to $\tau$\hsp:
\bel{hydro-1}
s(\tau)~=~\frac{s_0\tau_0}{\tau}~,
\ee
where $s_0=s(\tau_0)$. In this particular case Eqs. (\ref{eps-tau})
and (\ref{hydro-1}) are equivalent.
From Eqs.~\mbox{(\ref{eq:dSdy}) and (\ref{hydro-1})} one can see that $dS/d\eta$ is conserved
during the hydrodynamical expansion of~chemically equilibrated matter.
In a general case of a time-dependent $\lambda$ Eqs. (\ref{eps-tau})
and~(\ref{hydro-1}) are not equivalent, and the entropy $dS/d\eta$ is not conserved,
but increases during the hydrodynamic expansion (see below).

\noindent\textit{Freeze-out condition}\\
We assume that the Bjorken solution is valid until
the ''freeze-out'' (hyper)surface $\tau=\tau_f$.
Below we analyze a purely central Pb+Pb collisions at the LHC bombarding
energy of $\sqrt{s_{_{\rm NN}}}=2.76$~TeV. The freeze-out time $\tau_f$
will be determined by the condition
\bel{Tf}
T(\tau_f) = 15\hsp 6~{\rm MeV}.
\ee
Such a temperature value has been extracted~\cite{PBM} from
the thermal fit of hadron ratios observed in the considered reaction.

To get numerical estimates we use the approximate relation~{\cite{Hwa85}}
between the total entropy per unit space-time rapidity and
the rapidity density of pions
\bel{s0}
\frac{dS(\tau_f)}{d\eta}=\nu\,\left.\frac{dN_\pi}{dy}\right|_{{y=\eta}}=
\pi R^{\hsp 2}_A\hsp s\hsp(\tau_f)\hsp\tau_f\,,
\ee
where $\nu$ is the entropy per pion at the freeze-out stage of a heavy-ion collision. Note
that commonly used value $\nu=3.6$~\cite{shuryak} does not take
into account that a~large part of entropy is carried by
heavy mesons ($\rho, \omega \ldots$) and baryon-antibaryon pairs~($N,\ov{N},\Delta,\ov{\Delta} \ldots$).
The decay of hadronic resonances gives a significant fraction of observed pions.
Our calculations within the hadron resonance gas model~\cite{Sat13,VAG15} shows that $\nu\simeq 6.3$ at
$T\simeq 156~\textrm{MeV}$ and vanishing net baryon density.
Using experimental data of Ref.~\cite{ALICE13} we obtain that
\mbox{$\left. dN_\pi/dy\right|_{y=0}\simeq 2700$}\hsp . Substituting $R_A\,{\simeq}\,6.5$~fm
and $\nu = 6.3$ into \re{s0} we get the estimate
\bel{tenes}
\frac{dS(\tau_f)}{d\eta}\simeq 1.7\cdot 10^4,
\ee
which is used in our numerical calculations~(see next section).

\section{Numerical Results}

We compare two scenarios: the equilibrium QGP with the quark fugacity ${\lambda}=1$
and the chemically nonequilibrium uQGP
with ${\lambda=\lambda(\tau)<1}$. In the second scenario we assume that
${\lambda} (\tau_0)\ll 1$ and $\lambda\rightarrow 1$ at later times.
Below we use the parametrization
\bel{lambda}
\lambda\hsp (\tau)~=~1~-~\exp\left(\hsp\frac{\tau_0-\tau}{{\tau_*}}\right),
\ee
where $\tau_*$ is the model parameter characterizing the  {quark chemical} equilibration time.
Calculations of different authors gives different estimates for $\tau_*$
ranging from \mbox{$\tau_*\sim 1~\textrm{fm}/c$}~\cite{Ruggieri2015} to $\tau_*\sim 5~\textrm{fm}/c$~\cite{XuGreiner}.
One should have in mind that this parameter may depend on the combination of nuclei and the bombarding
energy. We expect that $\tau_*$ will be larger for peripheral events and lighter combinations of nuclei.
Figure~\ref{fig:La} shows the time dependence of $\lambda$ for several values of the parameters
$\tau_0$ and $\tau_*$. At small initial times, $\lambda (\tau)$ is only slightly sensitive to~$\tau_0$.
The chemically equilibrated case ($\lambda=1$) can be obtained at $\tau_*\to 0$. The case of a~pure glue
plasma corresponds to the limit $\tau_*\to\infty$.
\vspace*{3mm}
\begin{figure*}[hbt!]
\centering
\includegraphics[width=0.49\textwidth]{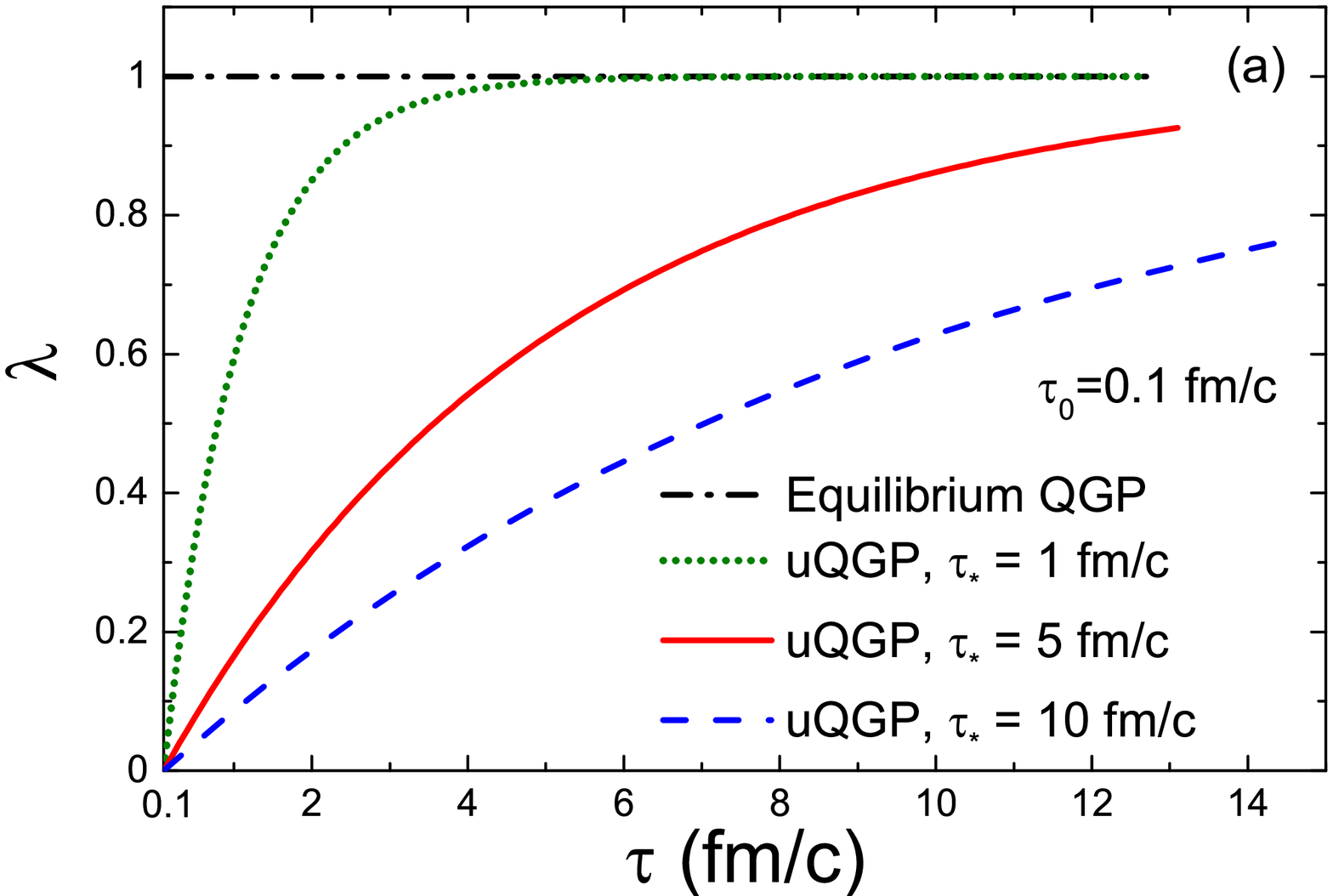}
\includegraphics[width=0.49\textwidth]{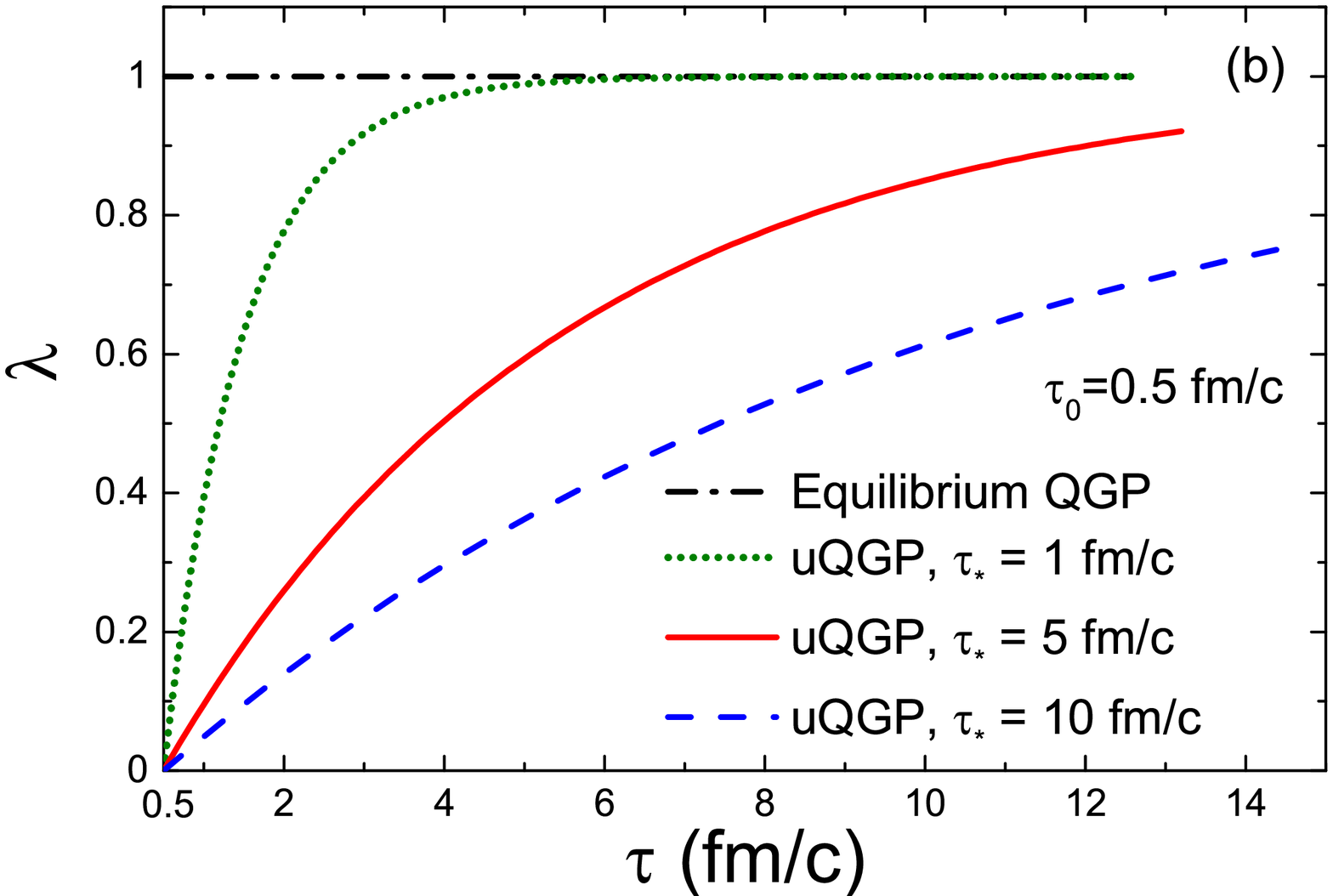}
\caption[]{
The  {quark fugacity} $\lambda$ as a function of proper time $\tau$ for two evolution scenarios
with (a) $\tau_0 = 0.1~\textrm{fm}/c$ and (b) $\tau_0 = 0.5~\textrm{fm}/c$\hsp . The dotted, solid and dashed lines correspond to uQGP, with parameters $\tau_* = 1~\textrm{fm}/c,~ 5~\textrm{fm}/c$ and $10~\textrm{fm}/c$\hsp, {respectively}.
}\label{fig:Lambda1}
\end{figure*}

Let us consider in more details the dynamics of the uQGP in the Bjorken model.
Using Eqs.~(\ref{eps-tau}) and (\ref{toten}), one gets the relations
$\varepsilon\propto T^4(1+\alpha\Lambda_4)\propto \tau^{-4/3}$. This gives the following equation
for temperature at \mbox{$\tau\geqslant\tau_0$}
\bel{tau-tem}
T=T_0\left(\frac{\tau_0}{\tau}\right)^{1/3}\big(1+\alpha\Lambda_4\big)^{-1/4}.
\ee
The explicit expression for $\Lambda_4 (\tau)$ is obtained by substituting (\ref{lambda}) into~\re{fdi4} with \mbox{$n=4$}\hsp .
As compared to the evolution of uQGP in the static box [see~\re{temfe}], the temperature contains the additional factor $(\tau_0/\tau)^{1/3}$. A stronger cooling in the expanding plasma occurs due to the work of pressure gradients
which accelerate fluid elements in the Bjorken model.

It is useful to rewrite~\re{tots} in the form
\bel{tots2}
s=s_0\left(\frac{T}{T_0}\right)^3\big(1+\alpha\Lambda_4-\beta\Lambda_3\ln{\lambda}\big),
\ee
where $s_0$ is the initial value of the entropy density. The latter is given by the first factor
in the right hand side (r.h.s.) of (\ref{tots}) taken at $T=T_0$\hsp . Using~Eqs.~(\ref{eq:dSdy}),
(\ref{tau-tem}), and (\ref{tots2}) we get the~equation for the total entropy of the QGP per unit space-time rapidity
\bel{dsde1}
\frac{dS(\tau)}{d\eta}=\frac{dS(\tau_0)}{d\eta}\cdot\frac{1+\alpha\Lambda_4-\beta\Lambda_3\ln{\lambda}}
{(1+\alpha\Lambda_4)^{\hsp 3/4}}\,.
\ee
The first factor in the r.h.s. is given by~\re{eq:dSdy} with the replacement $s\tau\to s_0\tau_0$. As seen
from the comparison with~\re{ss0ef}, we get the same entropy enhancement factor (as a function of $\lambda$)
as for the box equilibration in the fixed-energy case.
\begin{table}
\begin{center}
\begin{tabular}{|c|c|c|c|c|c|l|c|}
\hline
$\hsp\tau_0\hsp (\textrm{fm}/c)$ & $\hsp\tau_*\hsp (\textrm{fm}/c)$  & $\hsp\tau_f^{\rm eq}\hsp (\textrm{fm}/c)$\hsp
&\hsp $T_0^{\hsp\rm eq}\hsp (\textrm{MeV})$& $\hsp\tau_f\hsp (\textrm{fm}/c)$\hsp &\hsp $T_0\hsp (\textrm{MeV})$&
~~~$\lambda_f$\hsp &\hsp {$\mu_f\hsp (\textrm{MeV})$} \\
\hline
0.1 & 1  & 12\hspm.7 & 779 & 12\hspm.7  & 1022 &~{1.00}~&0\\
0.1 & 5  & 12\hspm.7 & 779 & 13\hspm.2  & 1023 &~0.927~&$-12$\\
0.1 & 10 & 12\hspm.7 & 779 & 14\hspm.5  & 1024 &~0.763~&$-42$\\
0.5 & 1  & 12\hspm.7 & 456 & 12\hspm.7  & 598  &~{1.00}~&0\\
0.5 & 5  & 12\hspm.7 & 456 & 13\hspm.2  & 598  &~0.921~&$-13$\\
0.5 & 10 & 12\hspm.7 & 456 & 14\hspm.6  & 599  &~0.756~&$-44$\\
\hline
\end{tabular}
\end{center}
\caption{\label{tab:Ttau}
The values of initial temperature $T_0$, freeze-out proper time $\tau_f$, {quark
fugacity~$\lambda$ and quark chemical potential $\mu$ at $\tau=\tau_f$}  for all considered cases.}
\end{table}

 Substituting $\tau=\tau_f$ into (\ref{tau-tem}) and (\ref{dsde1}) and using Eqs.~(\ref{Tf}) and (\ref{tenes}) gives two coupled
equations for determining the initial temperature $T_0$ and the freeze-out time $\tau_f$. The results of their calculation
for several values of $\tau_0$ and $\tau_*$ are given below (we take the same parameters as in Fig.~\ref{fig:Lambda1} and consider
the number of flavours $N_f=3$). For comparison we also make calculations within the chemically equilibrated scenario\hsp\footnote
{
In this case we take same values of $\tau_0$, same pion multiplicity and temperature at freeze-out as for uQGP.
}.
The values of $T_0, \tau_f$ calculated for all considered combinations of parameters are shown in Table~\ref{tab:Ttau}.
The last two columns give the fugacity and the chemical potential of quarks at freeze-out. One can see that in all cases
the initial temperature significantly exceeds the equilibrium value.

\begin{figure*}[hbt!]
\centering
\includegraphics[width=0.7\textwidth]{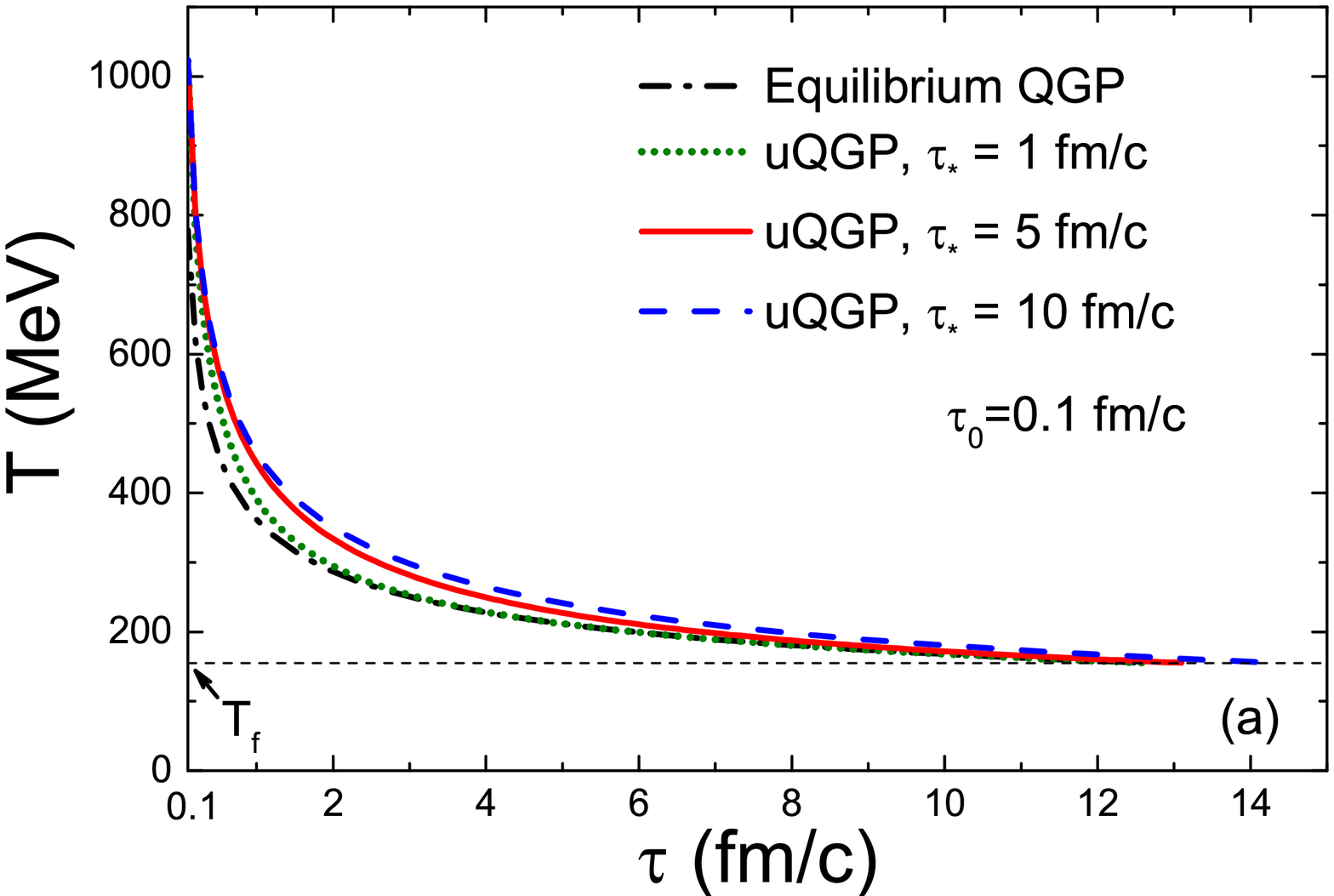}
\includegraphics[width=0.7\textwidth]{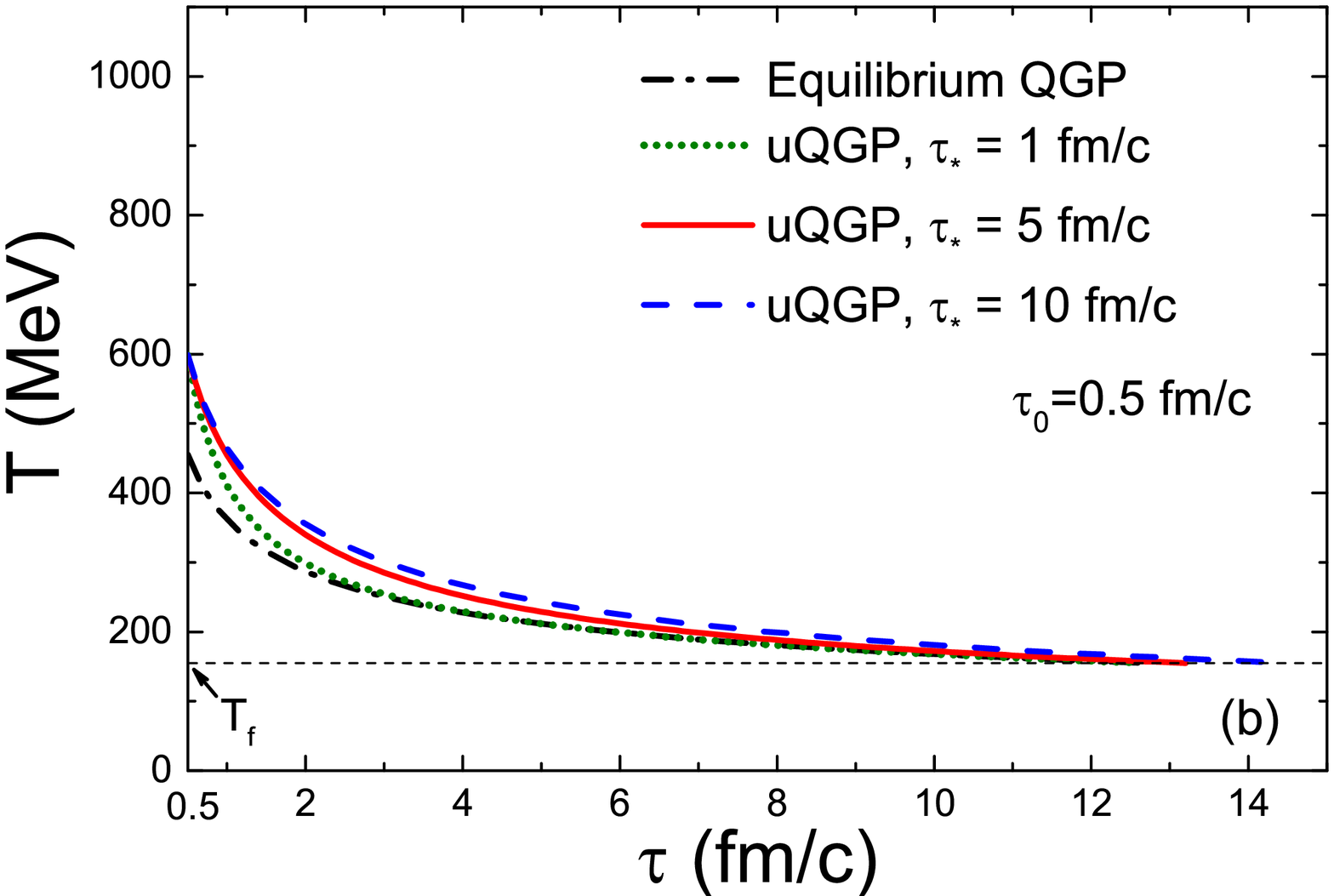}
\caption[]{
Same as Fig.~\ref{fig:Lambda1}, but for temperature as a function of $\tau$.
}\label{fig:Ttau1}
\end{figure*}

\begin{figure*}[hbt!]
\centering
\includegraphics[width=0.75\textwidth]{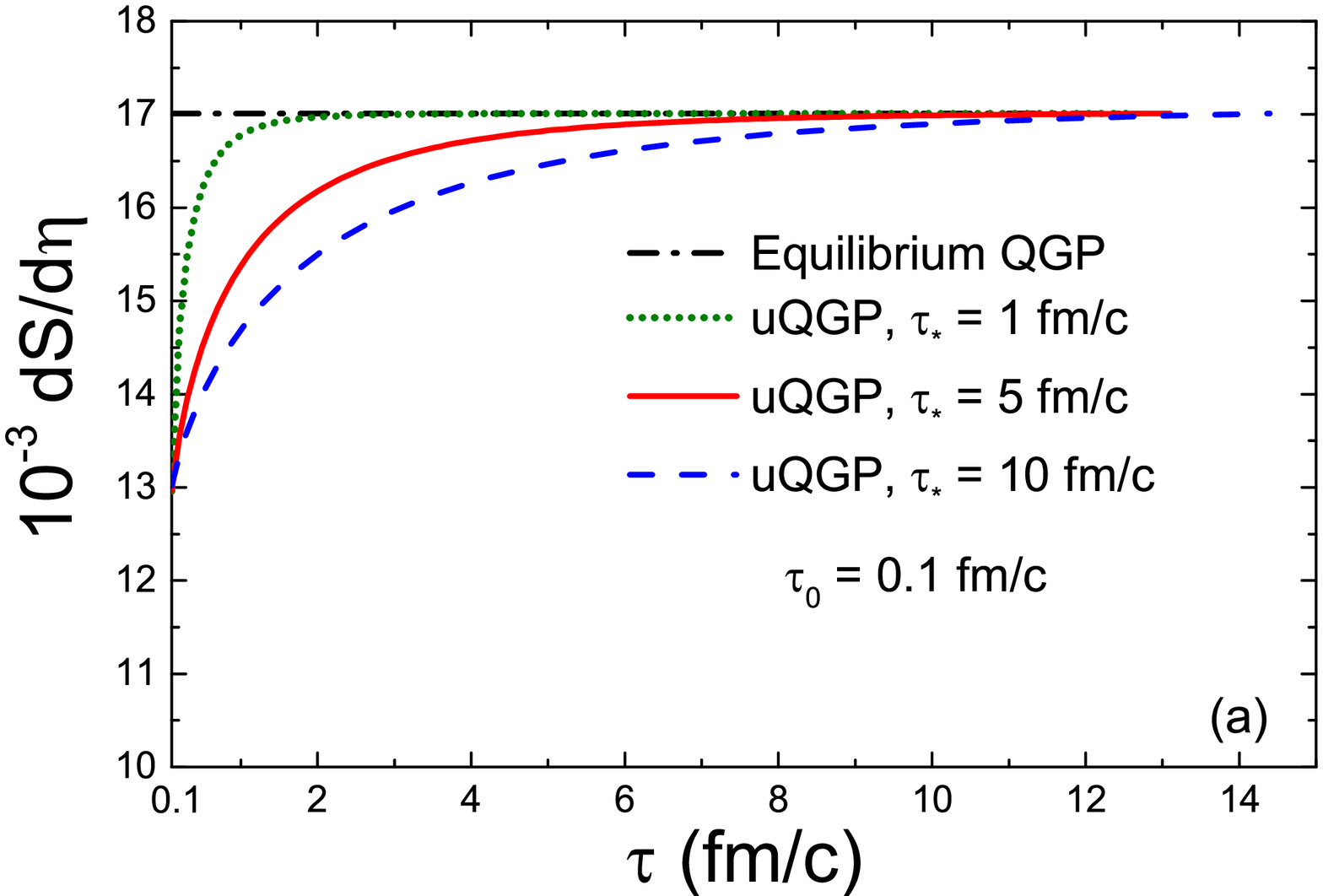}
\includegraphics[width=0.75\textwidth]{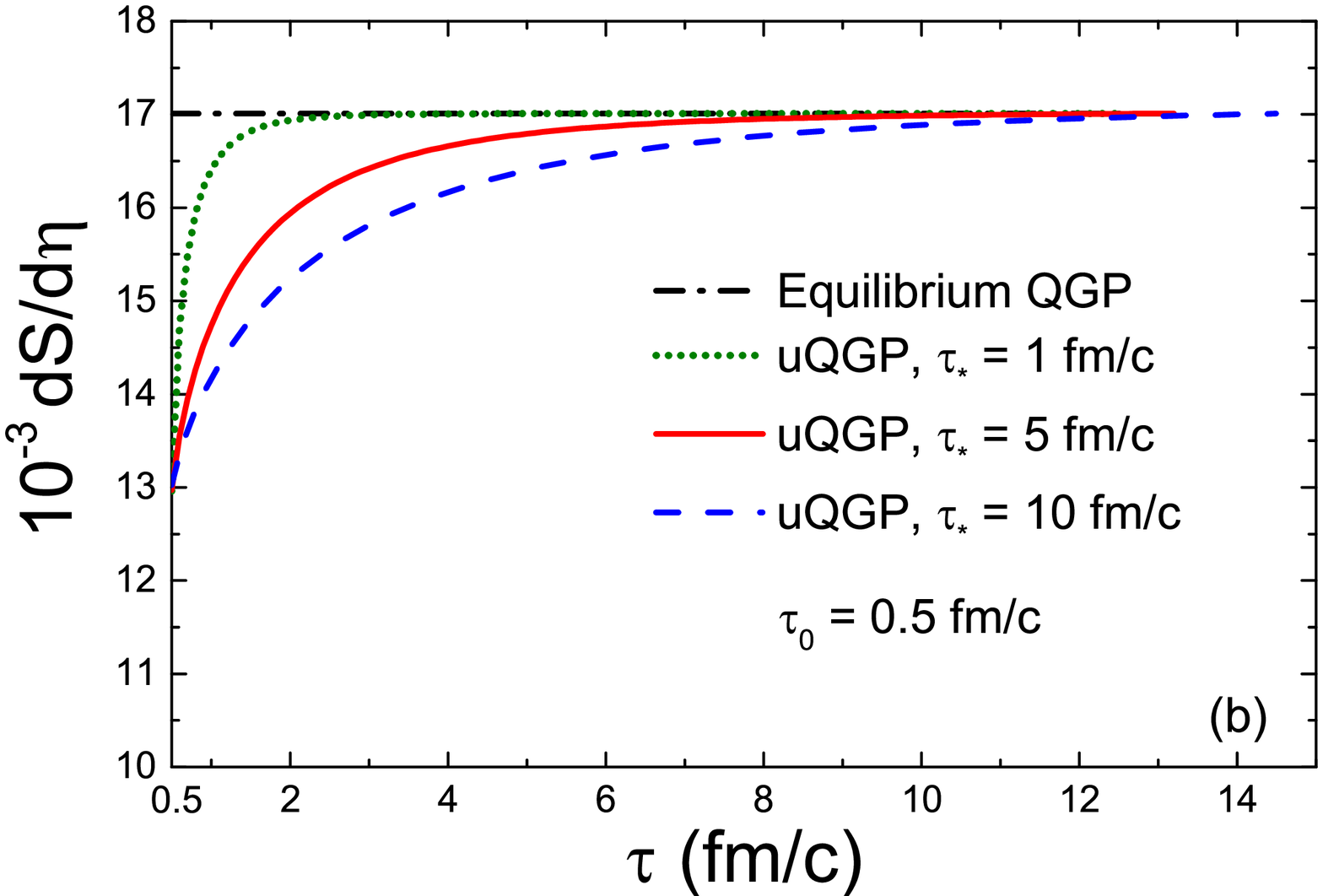}
\caption[]{
Same as Fig.~\ref{fig:Lambda1}, but for entropy per unit space-time rapidity.
}\label{fig:dSdy1}
\end{figure*}

A more detailed information is contained in Figs.~\ref{fig:Ttau1} and \ref{fig:dSdy1}.
The time dependence of temperature calculated from~\re{tau-tem} is shown in Fig.~\ref{fig:Ttau1}.
One can see that the deviation from equilibrium is most significant at the early stage,
and the pure glue initial scenario  predicts a higher temperature at any $\tau$.
Consequently, while there is a smaller amount of quarks during the
evolution of the uQGP than in the equilibrium case, they are generally hotter.
Note that a~two-fold increase of the equilibration parameter $\tau_*$,  from 5 to 10 fm/$c$,
only slightly changes the~cooling law $T=T(\tau)$  {of the undersaturated matter}.

The {evolution of the total} entropy per unit space-time rapidity calculated using~\re{dsde1}
is shown in Fig.~\ref{fig:dSdy1}. One can see that
the entropy in the chemically nonequilibrium scenario increases  {gradually}  with time and about~$25$\%
of  the final value (\ref{tenes}) is generated during the hydrodynamic expansion.
Our calculations show that the characteristic time of the entropy increase is of the order of $\tau_*$\hsp .
However, the total amount of produced entropy only weakly changes with $\tau_*$\hsp . This is also seen in Fig.~\ref{fig:entref1} where we show points A, B, and C (these points correspond, respectively, to the values of $\lambda_f$
from the first three lines of Table~I).

As seen from Fig.~\ref{fig:Lambda1} and Table~\ref{tab:Ttau}, the quark fugacity $\lambda$ remains smaller than
unity at the freeze-out hypersurface $\tau=\tau_f$ in the chemically nonequilibrium scenario. This implies the
suppression of quarks and antiquarks as compared to the equilibrium QGP even at the hadronization stage.
Such a behavior may influence the hadron composition measured in~central Pb+Pb collisions at LHC energies.
Note that the hadron resonance gas model~\mbox{{\cite{Cle06,And06,And09}}} can not explain the observed chemical composition of hadrons observed~\cite{p-pi} in these collisions.  In particular, the measured (anti)proton-to-pion ratios  are noticeably smaller than their equilibrium values for baryon-free matter. Calculations which allow deviations from the hadron equilibrium scenario are, thus, necessary.

Several theoretical models have been suggested to explain deviations from chemical equilibrium observed at LHC. Baryon suppression due  to inelastic collisions of hadrons at post freeze-out stage of a nuclear collision has been studied in \cite{becattini}. Possibility of pion enhancement due to positive pion chemical potentials was also investigated~\cite{begun}. In the present paper we propose an alternative explanation
of the observed suppression of the $p/\pi$ and $\ov{p}/\pi$ ratios. Indeed, according to the constituent quark structure of protons, antiprotons, and pions, one can estimate the suppression factor for both these ratios as $\lambda_f^3/\lambda_f^2=\lambda_f$. The latter can be noticeably below unity in undersaturated matter (see the corresponding column of Table~I). The same estimate can be obtained in the parton recombination model~\cite{Bas03} with chemically nonequilibrium effects as well as in the~statistical hadronization approach~\cite{Raf13}. The comparison of RHIC and LHC data shows~\cite{p-pi} that (anti)baryon-to-pion ratios are less suppressed at RHIC. This may be caused by a slower evolution of the fireball at lower
incident energies.

Another mechanism was proposed in Refs.~\cite{Sat13,Pra14}, where the $p/\pi$ suppression is
explained by annihilation of baryon-antibaryon pairs in dense hadronic matter created in nuclear collisions.
According to~Ref.~\cite{Sat13}, the $p/\pi$ ratios observed in central collisions can be reproduced if the annihilation persists until the temperature drops to $100-120$~MeV. Due to faster expansion and cooling of matter in peripheral events, one can expect stronger annihilation effects in more central collisions. However, the ALICE data~\cite{p-pi} reveal only small variations of the $p/\pi$ ratio as a function of centrality. This~discrepancy might be resolved by assuming some initial undersaturation of baryon-to-meson ratios, which increases with impact parameter\hsp\footnote
{
It is interesting that underpopulation of (anti)baryons at the posthadronization stage of nuclear collisions has been considered in Refs.~\cite{Nor10,Nor14}  within a model which takes into account the production and decay of Hagedorn resonances.
}. The latter assumption is rather natural because of reduced lifetimes of the deconfined phase in more peripheral events. We plan to extend the approach developed in~Refs.~\cite{Sat13,Pra14} for chemically nonequilibrium initial states in nuclear collisions.

\section{Conclusions and outlook}

We have investigated the dynamical evolution of deconfined matter with changing
chemical composition as expected in heavy-ion collisions at LHC energies. Two scenarios have been considered in details. The first one assumes that initially the system is composed exclusively of gluons, and later on quark-antiquark pairs are created during the characteristic time of $1-5~\textrm{fm}/c$. The second scenario assumes that the equilibrated QGP exists already at the initial stage.
The model parameters are chosen in such a way that the final pion multiplicity in both cases is equal
to the observed value for central Pb+Pb collisions. We predict that in the non-equilibrium scenario about
25\%
of final entropy is generated due to chemical equilibration
of plasma. We want to stress that this effect of entropy production is present in ideal hydrodynamics and it is attributed to increasing number of degrees of freedom. This is different from the case of entropy production via dissipative processes which are determined by transport coefficients and usually modeled by the viscous hydrodynamics. Obviously, the inclusion of chemically nonequilibrium effects may require
modification of the viscosity coefficients extracted from the fit of collective flow
observables~\cite{Rom07,Nie11}.

In the present work we do not develop any full-fledged formalism to describe the pure glue initial scenario.
Ideally, one would like to determine the dynamical evolution of matter created in heavy-ion collisions directly from experimental data. That would require using a model working in a reverse way, starting from the measured data, such as identified particle momentum spectra, and then proceeding backwards in time. Reversing a dynamical evolution in an ideal (1+1) dimensional  hydrodynamics has been performed in Ref.~\cite{InverseHydro}. Note, however, that
uncertainties in the final state measured in detectors increase when going backwards in time.
Furthermore, irreversible processes, associated with viscosity, as well as with particle production out of equilibrium, as discussed in this paper, lead to~an~increase of entropy
and further reduce the accuracy of this backtracing procedure.

A crucial test of the pure glue initial scenario may be provided by the electromagnetic probes, i.e.,
by emission of thermal photons and dileptons. This study will be presented
in a~forthcoming publication (our preliminary results are given in Ref.~\cite{Sto15b}). We also
plan to~perform a~more realistic calculation within
a (3+1) dimensional hydrodynamic model which takes into account the transverse motion of matter. Then
one can analyze the sensitivity of photon and hadron observables
to chemically nonequilibrium effects at early stages, and to~violation of Bjorken scaling at later
stages of a heavy-ion collision. The calculations can be made even more realistic
by introducing additional rate equations describing the~space-time evolution
of quark and gluon densities (see e.g.~\mbox{\cite{Bir93,Ell00,Mon14}}).

We are also going to study in more detail dynamics of the first order phase transition
as predicted in the pure gluodynamics. In particular, it will be interesting to study influence of this phase transition on flow observables (see~Refs.~\cite{Mer11,Iva15})\hspm . Another interesting possibility is supercooling/overheating  processes associated with the  {deconfinement phase transition}~\mbox{\cite{Herold,Steinheimer}}.

\begin{acknowledgments}
The authors thank C. Greiner, P. Huovinen and H. Niemi for valuable comments.
This work was partially supported by the Helmholtz International Center for
FAIR, Germany and by the Program of Fundamental Research
of the Department of Physics and Astronomy of National Academy of
Sciences of Ukraine.
L.M.S. and I.N.M. acknowledge a partial support from
the grant NSH-932.2014.2 of the Ministry of Education and Science
of the Russian Federation.
\end{acknowledgments}

\end{document}